\documentclass[
twocolumn,
a4paper
showpacs,
preprintnumbers,
amsmath,
amssymb,
aps,
]{revtex4-1}

\usepackage{graphicx}
\usepackage{dcolumn}
\usepackage{bm}
\usepackage{hyperref}
\usepackage{upgreek}
\usepackage{setspace}
\usepackage[super]{nth}

\begin{document}


\title{One-dimensional Si chains embedded in Pt(111)\\ and protected by a hexagonal boron-nitride monolayer}

\author{Silke \surname{Rose$^1$}}
\email{rose@physik.rwth-aachen.de}
\author{Peter \surname{Nemes-Incze$^{1,2}$}}
\author{Marco \surname{Pratzer$^1$}}
\author{Vasile \surname{Caciuc$^3$}}
\author{Nicolae \surname{Atodiresei$^3$}}
\author{Markus \surname{Morgenstern$^1$}}

\affiliation{$^1$II. Physikalisches Institut B and JARA-FIT, RWTH Aachen University, Otto-Blumenthal-Stra{\ss}e, 52074 Aachen}
\affiliation{$^2$Centre for Energy Research, Institute of Technical Physics and Materials Science,
1121 Budapest, Hungary}
\affiliation{$^3$Peter Gr\"{u}nberg Institut and Institute for Advanced Simulation, Forschungszentrum J\"{u}lich and JARA, D-52425
J\"{u}lich, Germany}

\date{\today}

\begin{abstract}
Using scanning tunneling microscopy, we show that Si deposition on Pt(111) at 300K leads to a network of one-dimensional Si chains. On the bare Pt(111) surface, the chains, embedded into the Pt surface, are orientated along the $\langle \overline{1} \overline{1} 2 \rangle$-direction. They disappear within a few hours in ultrahigh vacuum due to the presence of residual gas. Exposing the chains to different gases deliberately reveals that CO is largely responsible for the disappearance of the chains. The chains can be stabilized by a monolayer of hexagonal boron nitride, which is deposited prior to the Si deposition. The resulting Si chains are rotated by 30$^\circ$ with respect to the chains on the bare Pt(111) surface and survive even an exposure to air for 10 minutes.
\end{abstract}

\pacs{Valid PACS appear here}
\maketitle



\section{\label{sec:Intro}Introduction}

It is well known that surfaces and interfaces are distinct from their bulk counterparts due to the modified bonding configurations \cite{Zangwill1988}. A conceptually simple difference is an increased atomic density of the surface, which compensates the reduced charge carrier density from the missing atoms on top. This concept is realized, e.g., in the famous Au(111) herringbone reconstruction, where the additional atoms form a network of partial dislocation lines \cite{Barth1990}. A similar network of dislocation lines has been found on the Pt(111) surface, but there it is thermodynamically stable only at a sufficiently large Pt pressure in the gas phase \cite{Bott1993,Hohage1995} or at high temperatures \cite{Sandy1992}. A similar densification of the surface via dislocation lines is also found, if foreign atoms are offered to Pt(111) such as Co \cite{Lundgren1999,Grutter1994}, Cu \cite{Holst1998} or Cr \cite{Zhang1999}. This enables an elegant pathway to establish one dimensional (1D) structures of a certain chemical species, embedded in a chemically different matrix, via self-organization.

On the other hand, the driving force of the densification can be mitigated by adsorbates on top, which provide additional electron density. Indeed, lifting of surface dislocation networks has been observed for several adsorbates \cite{Michely1996,Tempas2016,Barth1996,Anic2016}. Consequently, the created 1D structures are typically quite fragile even under ultrahigh vacuum (UHV) conditions.

Here, we investigate a novel system belonging to the class of densified surfaces via dislocation lines, namely Si at the Pt(111) surface. It has been overlooked so far, probably due to its sensitivity to adsorbates. It is formed by a submonolayer deposition of Si on Pt(111) at room temperature, where a network of one-dimensional Si chains embedded into the Pt surface appears. The chain network continuously decays and is completely lifted after 4-10 hours in an ultrahigh vacuum (UHV) environment. We show that CO triggers the lifting of the chain network rather effectively, such that lifting is completed by a coverage of about one Langmuir. 
In order to protect the Si chains against CO and other adsorbates, we covered Pt(111) with a single layer of hexagonal boron nitride (h-BN) and offered the Si subsequently. This also leads to a network of embedded Si chains in Pt, but this network remains stable for days in UHV and can even be exposed to air for 10 min without a complete lifting. Surprisingly, the Si chains below the h-BN are oriented along the close-packed $\langle 1 \overline{1} 0 \rangle$-direction, i.e. they are rotated by 30$^\circ$ with respect to the Si chains on bare Pt(111). We speculate that this rotation is due to an anisotropy in the mechanical properties of the h-BN overlayer.

\section{\label{sec:Exp}Experimental}
Experiments were performed in an UHV system with a base pressure of $1\cdot 10 ^{-8}\,\rm{Pa}$ maintained by an ion getter pump and a titanium sublimation pump. The system is equipped with a sputtering ion gun, a silicon evaporator, an e-beam heater, a resistive heating station, a low energy electron diffraction (LEED) system, various gas inlets and a home-built room temperature scanning tunneling microscope (STM) \cite{Geringer2009}.
The Pt single-crystal was cleaned by several cycles of argon ion sputtering at 720\,K, heating to 870\,K in an O$_2$ atmosphere ($p=3\cdot 10 ^{-5}\,\rm{Pa}$) and flash annealing to 1270\,K \cite{MUSKET1982143, Morgenstern1997}. Cleanliness was checked by LEED and STM.
Hexagonal boron nitride monolayers were grown subsequently by a self-limiting epitaxial process \cite{CAVAR20081722,PREOBRAJENSKI2007119,Preobrajenski2007}. For this, the Pt crystal was heated to 1100\,K and exposed to borazine from an inlet directly above the sample, resulting in a UHV chamber pressure of $1\cdot 10 ^{-6}\,\rm{Pa}$.
Si was evaporated at room temperature directly onto the Pt(111) surface or after h-BN coverage both at a rate of 0.01\,monolayers/min employing an electron beam evaporator with a tantalum crucible \cite{MYSLIVECEK2002193}. Cleanliness of the deposited silicon was checked by reproducing the silicene 4$\times$4 reconstruction on Ag(111) \cite{Vogt2012, Jamgotchian2012}. This reconstruction was also used to calibrate the evaporator in terms of monolayer per min, where one monolayer (ML) refers to one Si atom per Pt atom in the top layer. 
In order to avoid radicals provided by a hot filament, the ionization gauge was turned off during all measurements on the bare Pt(111) surface. Pressures are then estimated from the current of the ion getter pump as previously calibrated by the ion gauge.
STM measurements were carried out at 300\,K using tungsten tips prepared by electrochemical etching and subsequent short heating inside the UHV chamber in order to remove the tungsten trioxide layer formed during the etching process \cite{Lucier2005}.
The voltage is applied to the sample and STM images are recorded in constant current mode at tunneling current $I$.

\section{\label{sec:Comp}Computational setup}
Our first-principles electronic structure calculations were performed using the density functional theory (DFT) \cite{Hohenberg1964, Kohn1965} and the projector augmented wave method (PAW) \cite{Blochl1994} as implemented in the VASP code \cite{Kresse1993, Kresse1996, Kresse1999}. To describe the van der Waals interactions present in the hBN/silicene/Pt(111) and silicene/hBN/Pt(111) systems the non-local correlation energy functional vdW-DF2 \cite{Lee2010} was employed together with a reoptimized \cite{Hamada2014} Becke (B86b) exchange energy functional \cite{Becke1986}. The h-BN/silicene and silicene/h-BN on Pt(111) were modeled by slabs containing six Pt atomic layers with an 5$\times$5 in-plane unit cell. The ground-state geometries and the corresponding electronic structure of these systems have been obtained for a kinetic energy cut-off of 450\,eV and a threshold value of the calculated Hellmann-Feynman forces of $\sim$0.01\,eV/\AA . Furthermore, the Brillouin zone integrations were performed using a 12$\times$12 k-mesh. 

\section{\label{sec:RaD}Results and Discussion}

\begin{figure}[!ht]
\includegraphics[width=0.49\textwidth]{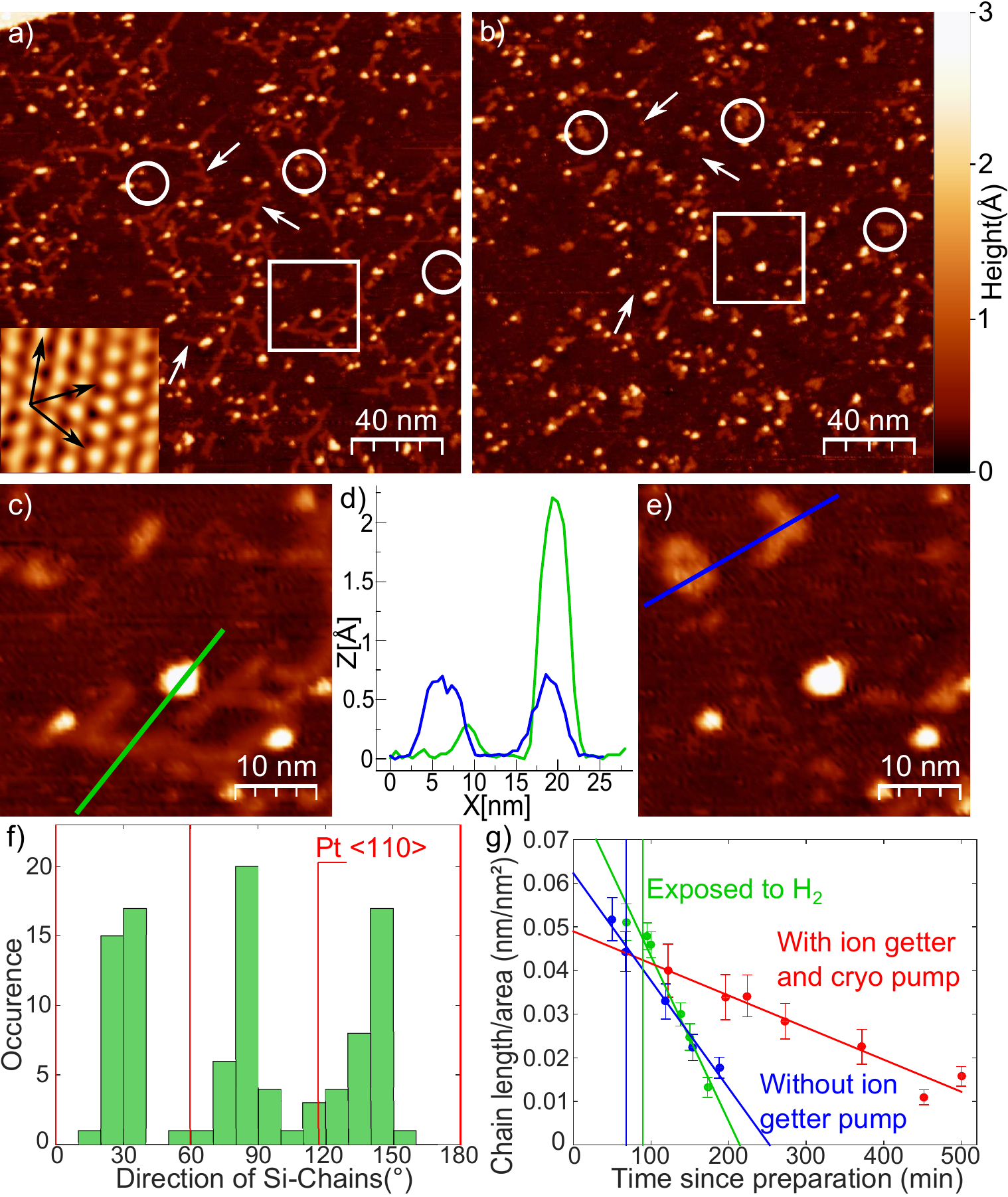}
\caption{\label{fig:Pt} (Color online) 1D Si chains embedded in Pt(111). (a) STM image of Pt(111) after evaporation of 0.03\,ML of Si at room temperature and additional waiting time of 50 min. $V$=0.3\,V; $I$=0.5\,nA. (b) STM image of roughly the same area as in (a) taken 140 minutes later. The ion getter pump was turned off after acquisition of the first image, which results in a continuous rise of the pressure to $9\cdot 10^{-6}\,\rm{Pa}$ within 4 hours. White arrows and white circles mark features which disappear and grow as the chain network is lifted, respectively. Inset in (a) shows atomic resolution on bare Pt(111) as used for the calibration of the directions. $\langle 1 \overline{1} 0 \rangle$-directions are marked. (c),(e) Cut outs as marked by the white rectangles in (a),(b), respectively, showing a disappearing chain network (lower part) and growing, more compact, embedded areas (upper left corner) in detail. (d) Profile lines as marked in (c),(e). (f) Histogram of the directions of the 1D chains (green) employing different STM images. Red, vertical lines mark the $\langle 1 \overline{1} 0 \rangle$-directions of Pt(111). (g) Development of the 1D chain coverage. The red points showcase a sequence where, in addition to the chamber's ion getter pump a cryo pump was operating for the final flash of the Pt(111) crystal, the Si deposition and during the STM measurements, thus reducing residual gas atom density. The blue data points were acquired while the ion getter pump was turned off, 66 minutes after preparation (marked by the vertical blue line). This results in a continuous rise of the pressure up to $9\cdot 10^{-6}\,\rm{Pa}$ within 4 hours. The green data points were acquired with the ion getter pump operating and additional H$_2$ exposure ($6\cdot 10^{-7}\,\rm{Pa}$) starting 88 minutes after preparation (marked by the vertical green line).}
\end{figure}

Figure \ref{fig:Pt}(a) shows the STM image of a clean Pt(111) surface, 50 minutes after exposure to 0.03\,ML Si. Small clusters about 200\,pm in height are visible. In addition, a network of faint lines is apparent (arrows). The lines are 25\,pm high and appear few nm in width, as shown in the profile line in Fig. \ref{fig:Pt}(d). Additionally, small extended areas of about 60-70\,pm in height are observed (circles and blue profile line in Fig. \ref{fig:Pt}(d)). 140 minutes later, an image of the same area (Fig. \ref{fig:Pt}(b)) reveals that most of the lines have disappeared (white arrows) while some of the 60-70\,pm high areas have grown in size (white circles). Figures \ref{fig:Pt}(c) and (e) display enlarged views of the same area (white squares in Figs. \ref{fig:Pt}(a) and (b)) highlighting the disappearance of the network of lines (lower half) and the growth of the  more compact areas (upper left corner). About four hours after preparation (corresponding to 300\,Langmuir of residual gas exposure) we do not find any of the lines anymore.

The lines exhibit a preferential orientation close to the $\langle\overline{1}\overline{1} 2 \rangle$ direction, as shown by the histogram in Fig. \ref{fig:Pt}(f). For the histogram, the angle of the line direction relative to the atomic rows (inset in (a)) was determined by visual inspection. We evaluated several images after three preparation cycles.

We assume that the lines are embedded in the Pt(111) surface for the following reasons. Since we observe different structures induced by the Si deposition, one having the typical height of an atomic step (2\,\AA), it is rather unlikely that the other structures are also located on top of the Pt(111) surface. Moreover, the lines have the same height as observed for dislocation lines on the Pt(111) surface (25\,pm) \cite{Bott1993, Hohage1995}. Hence we argue that the Si is embedded into the Pt(111) surface along the $\langle\overline{1}\overline{1} 2 \rangle$ direction. This is supported by the fact that no movement of the lines could be induced by the STM tip up to V=0.3\,V and I=9.5\,nA. We propose an atomistic model of the embedded Si atoms further below.

The speed at which the lines disappear depends on the amount and composition of the residual gas. Figure \ref{fig:Pt}(g) shows the development of the line lengths per area in different environments. The line length was determined by first flattening the STM images. We applied a \nth{3}-order plane-fit to remove the slight curvature of the images (caused by piezo creep) and subsequently a conditional line fit, which excludes protruding areas such as Si lines and clusters in order to prevent offsets due to the inhomogeneous distribution of those. Then, the $\sim$200\,pm high clusters were identified by their protruding height and removed from the image. As the clusters are broadened due to the convolution with the tip, the identified cluster areas were increased by 1\,nm in width to also remove the apparent clusters' rims. The area of the lines was then identified by using the image area exhibiting a height between the half-height of the lines and their full height, which was itself determined by averaging the height of 20 profile lines placed through chains by visual inspection. In order to finally get a length, the resulting area value was divided by the average line width, which is taken as the width of the line perpendicular to the $\langle 1\overline{1} 0 \rangle$ directions. The error of the resulting line length per area is dominated by the error of the determined line height, influencing the area value via the resulting cut-off at half-height.

Under optimized UHV conditions, at $p<1\cdot 10^{-8}\,\rm{Pa}$, which were obtained when an additional cryo pump is turned on for the final flash during the Pt cleaning and kept running throughout the STM measurement, some lines are still observed 8 hours after Si deposition, i.e, after exposure to 4\,Langmuir of residual gas (Fig. \ref{fig:Pt}(g)). 
Without the ion getter pump operating and without the cryo pump, the pressure is by more than a factor of 100 larger, while the slope of the line length/area vs. time is steeper by only a factor of 3-4 (Fig. \ref{fig:Pt}(g)). Hence, one suspects that only part of the residual gas is responsible for the lifting of the Si lines. 
Mass spectroscopy reveals that the amount of CH$_n$ increases by about three orders of magnitude and the amount of H rises by about one order of magnitude (mainly due to the rise of carbon hydroxides) while the amounts of H$_2$ and CO rise by a factor of about 1.5 only after turning off the ion getter pump. 
Thus, H$_2$ and carbon monoxide are likely candidates lifting the Si lines. To test this hypothesis, either H$_2$ or CO were deliberately let into the chamber during the STM measurements. In both cases, the samples were prepared while the ion getter pump was operating. After a first STM image to check the success of the  preparation, the gas was let in via a dosing valve. An exposure of $\sim$ 40\,Langmuir H$_2$ was necessary until the Si lines disappeared completely. This is 10 times more than the dose of residual gas required at the optimal UHV conditions. Hence, while the additional H$_2$ fastens the lifting of the lines, it cannot be solely responsible for their fragility under UHV conditions.

\begin{figure}[!ht]
\includegraphics[width=0.35\textwidth]{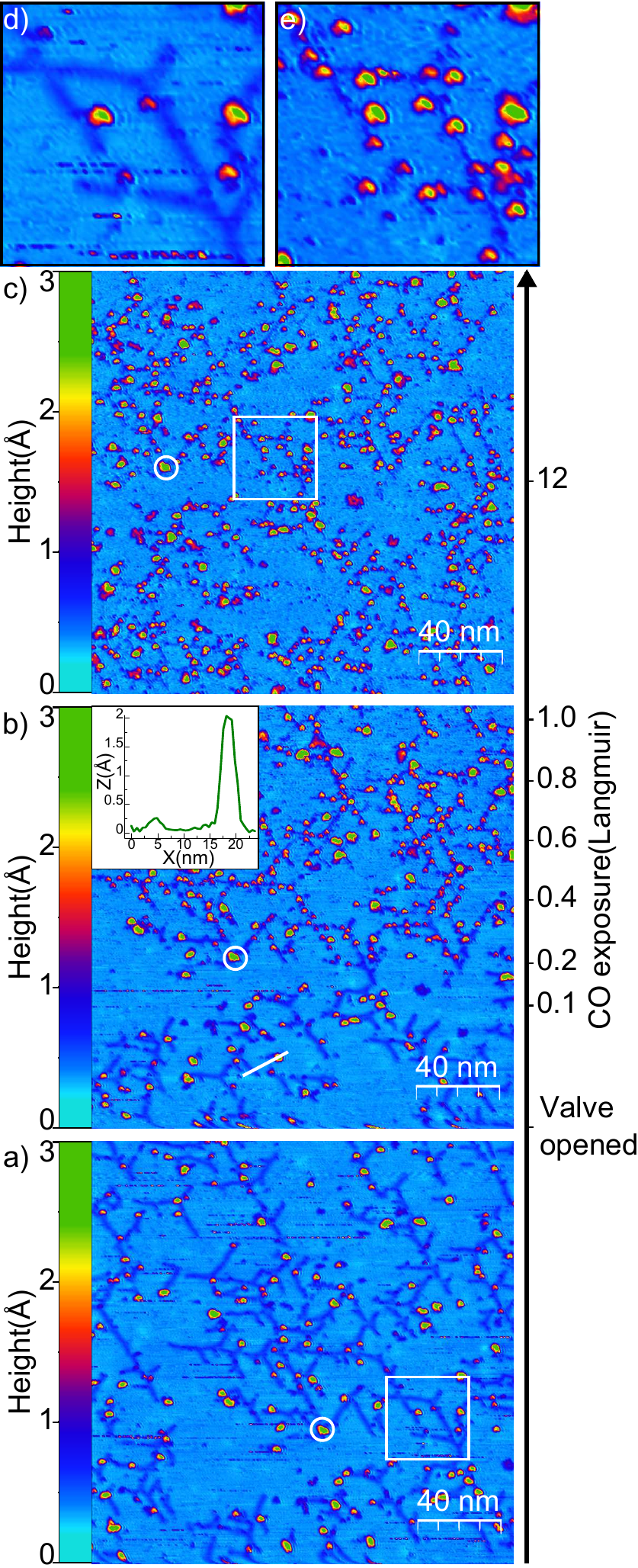}
\caption{\label{fig:CO} (Color online) Development of the 1D Si chains during CO exposure. (a) STM image of Pt(111) after evaporation of 0.07\,ML of Si, recorded 47 minutes after preparation. (b) STM image of the same area during exposure to CO ($7\cdot 10^{-7}\,\rm{Pa}$). The image is recorded from bottom to top and the CO exposure at a particular line is given on the right. Note that the scale is not linear due to a changing CO pressure. The inset shows the marked height profile. (c) Subsequent STM image of the same area during continuing CO exposure. All images were recorded at $V$=0.3\,V and $I$=0.5\,nA. (d),(e) Cut outs as marked by the white rectangles in (a),(c), respectively, showing the disappearance of the chains and growth of small clusters along them within the same surface area.}
\end{figure}

In contrast, about one Langmuir of CO is sufficient to reduce the line density significantly, as can be seen in Fig. \ref{fig:CO}. Figure \ref{fig:CO}(a) shows a STM image prior to gas exposure. The contrast is optimized to increase the visibility of the lines showing up in a dark blue color. At the bottom of Fig. \ref{fig:CO} (b), which is scanned from bottom to top, CO was let into the chamber. Over the course of that image, the number of lines start to diminish, while additional clusters appear on the surface. Figure \ref{fig:CO}(c) shows the same area again revealing the nearly complete absence of lines. Hence, CO is much more efficient in lifting the lines than H$_2$.
However, the resulting morphology is different than under UHV conditions. Exposed to residual gas, the disappearance of the lines mainly resulted in the growth of small embedded areas with a height of about 60-70\,pm, while the exposure to CO results in the growth of small clusters with height of $\sim$200\,pm along the former lines, while no embedded areas grow. In residual gas these embedded areas did not disappear even after 24\, hours and no growth of clusters was observed. Thus, other components of the residual gas might play a role in the exact process of lifting the lines, though CO plays a major role, likely initiating the process.

\begin{figure*}
\includegraphics[width=1\textwidth]{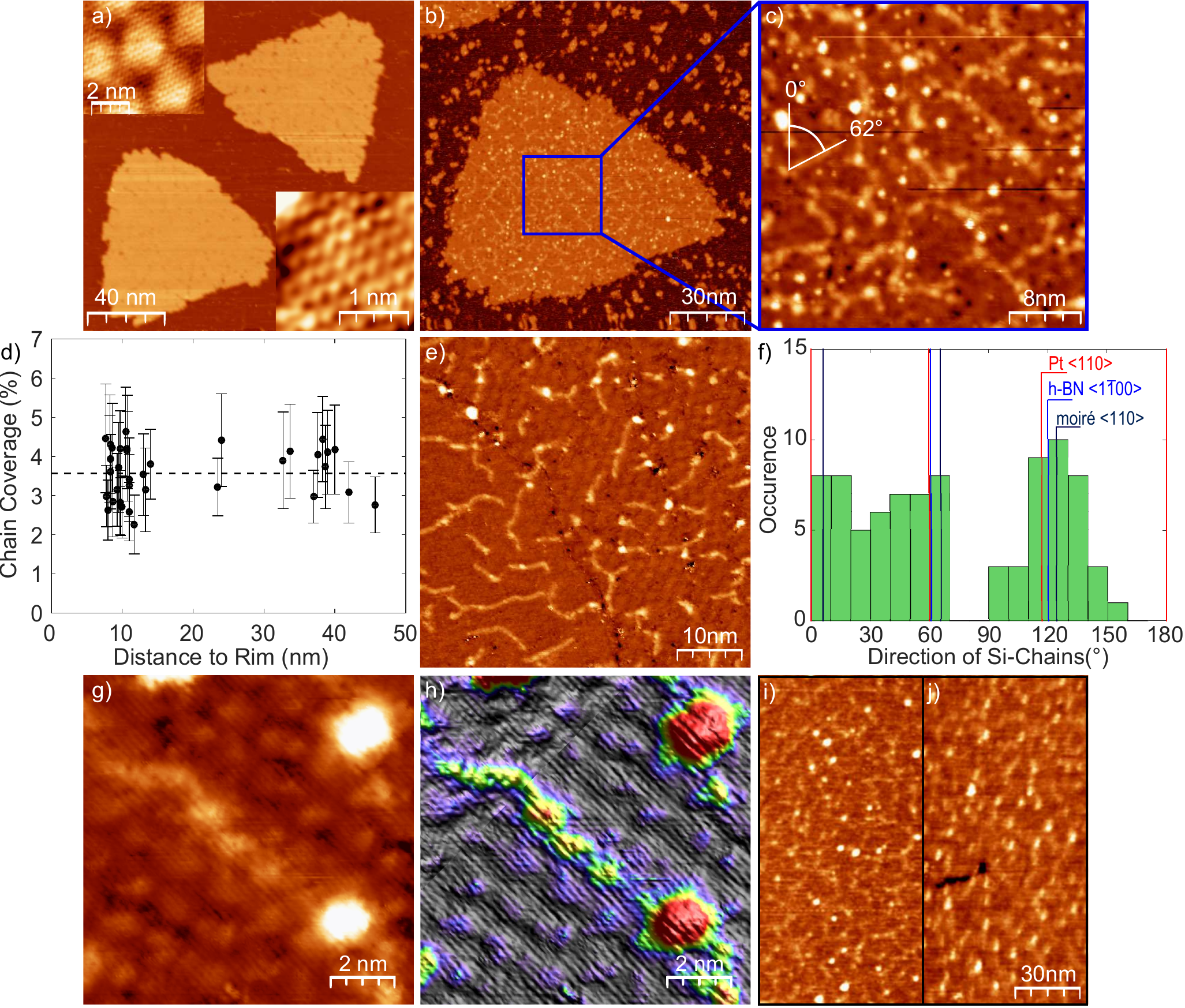}
\caption{\label{fig:BN} (Color online) 1D Si chains in Pt(111) protected by a monolayer h-BN. (a) STM image of h-BN islands grown on Pt (111). $V$=0.5\,V; $I$=0.1\,nA. The upper inset showcases the moir\'{e} pattern with atomic resolution of the h-BN layer ($V$=0.1\,V; $I$=2\,nA), the lower inset shows a zoom-in for better visibility of the atomic resolution. (b) STM image of a h-BN island on Pt(111) after deposition of 0.07\,ML Si $V$=-0.5\,V; $I$=0.2\,nA. (c) Cutout from (b) after additional \nth{2}-order plane fit. (d) Coverage with presumably one atom thick Si chains (below h-BN area) in dependence of the distance to the h-BN rim. (e) STM image of a sample covered completely with h-BN and subsequently covered with 0.07\,ML of Si. $V$=0.3\,V; $I$=0.8\,nA. The grain boundary (dark line) does not influence the density of the Si chains. (f) Histogram of the directions of the 1D chains (green). The data originate from chains below several different h-BN islands. Blue and red lines mark the $\langle 1\overline{1}00 \rangle$ and $\langle 1\overline{1}0 \rangle$-directions of the h-BN(0001) and Pt(111), respectively. The black line marks the observed $\langle 1\overline{1}0 \rangle$-direction of the hexagonal moir\'{e}, which is rotated by an angle of 5$^\circ$ with respect to the Pt$\langle 1\overline{1}0 \rangle$-direction. (g) Atomically resolved STM image of the h-BN layer covering a Si chain. $V$=50\,mV; $I$=3\,nA. A moir\'{e} with about 2\,nm length and atomic rows continue across the chain. Due to tip effects, the atoms are more clearly resolved in one direction. (h) 3D representation of the image in (g) viewed from the top. The color scale and illumination were chosen to emphasize the continuation of the atomic rows across the chain and the clusters. The upper left corner has the same color scale, but a slightly different illumination to visualize individual atoms. (i) STM image after another preparation identical to the one for (e). $V$=0.5\,V; $I$=0.5\,nA. (j) Same sample as in (i) after being taken out of UHV and exposed to air for 10 minutes. $V$=0.2\,V; $I$=1\,nA.}
\end{figure*}

Next, we describe the experiments employing the h-BN protection layer. Figure \ref{fig:BN}(a) shows Pt(111) exposed to borazine for 10 minutes in order to grow h-BN islands. The h-BN forms mostly triangular islands \cite{Hagen2016}. Afterwards the sample was exposed to 0.07\,ML of Si at room temperature. A STM image of an h-BN island after exposure to Si is shown in Fig. \ref{fig:BN}(b).  The bare Pt(111) surface surrounding the h-BN island is covered with clusters originating from the Si deposition without any remaining traces of lines. The zoom into the h-BN island in Fig. \ref{fig:BN}(c), however, shows one dimensional lines with some small clusters in between. No extended areas of sub-monolayer height were observed below h-BN islands. 

The first question to be addressed is whether the lines are located below or on top of the h-BN. We argue that they are located below, as supported by the following arguments. 

First, the atomic lattice of the h-BN as apparent in STM images continues over the lines and clusters. Figure \ref{fig:BN}(g) and (h) show the same atomically resolved STM image which includes a Si line in two different representations. The image was processed in the same way as described for Fig.\ref{fig:Pt}. No fast Fourier transform (FFT) filter has been applied to the images.The atomic corrugation of the h-BN atoms is only about 2\,pm, so it is rather hard to see them in Fig. \ref{fig:BN}(g) at normal color code representation. Optimizing the contrast in Fig. \ref{fig:BN}(h), however, clearly reveals the continuation of the atomic rows of the h-BN across the Si lines and clusters. This can only be rationalized if the lines and clusters are below the h-BN layer.

Second, our DFT calculations of a silicene sheet between h-BN and Pt(111) reveal a weak physisorption between Si and h-BN (91\,meV per atom), but a stronger chemisorption of the Si to the Pt (1.435\,eV). Thus, from an energetical point of view, it is much more favorable for the Si to bind to the Pt, implying that the Si atoms gain energy by moving below the h-BN cover. 

Third, no movement of the clusters and lines could be induced by the STM tip down to V=25\,mV and up to I=26\,nA. Our density functional theory (DFT) calculations of a silicene sheet on h-BN on Pt(111) reveal that the silicene sheet is only physisorbed with an energy gain of 127\,meV per atom on the h-BN surface. Consequently, the Si atoms, if only attached to the h-BN, should even be mobile at 300\,K, though dangling bonds from the Si clusters, might also prevent their mobility on the h-BN. 

Finally, contrary to the lines on the bare Pt(111), the lines below the h-BN are long-term stable in UHV. They are still observed nine days after the preparation and even largely withstand an exposure to air for about 10 minutes corresponding to $6\cdot 10^{11}$ Langmuir. Figure \ref{fig:BN}(i) and (j) showcase the sample before and after air exposure. The area surrounding some defects appears darker after the exposure probably due to oxidation, but lines are still observable in the STM image. Considering the instability of the lines on bare Pt(111), this points to an excellent protection below the monolayer of h-BN.

Thus, we conclude that the Si is intercalated between the h-BN and the Pt. By analogy, we assume that the lines being about 25\,pm in height are embedded into the Pt(111) surface, while the $\sim$200\,pm high clusters are located on top of the Pt(111) surface. The driving force for the embedding is likely again the increase of electron density in the Pt(111) surface, which does not profit significantly from the h-BN due to its large band gap of 4.5\,eV \cite{Bhattacharya2012}.

It is remarkable that Si can be implanted below h-BN from a thermal evaporator. Monolayer h-BN has been shown to protect other intercalated structures. For example, Ar$^{+}$ and Xe$^{+}$ ions have been implanted below h-BN layers on metal surfaces forming gas-filled blisters \cite{Valerius2017, Cun2013, Cun2014, Cun2014a}. Trapped Xe$^{+}$ ions have been observed even after annealing up to 1550\,K, at which point the h-BN decomposes freeing the Xe$^{+}$ \cite{Valerius2017}. Ar$^{+}$-filled blisters survive exposure to air \cite{Cun2014a}. However, these structures were created by bombarding the surface with low to high-energy ions, as the expected displacement threshold energies for B and N are $\sim$20\,eV \cite{Kotakoski2010}. The thermal energy provided to the Si atoms is small in comparison.

In order to determine the pathway of Si intercalation, we analyzed the coverage with lines as a function of distance to the rims of the h-BN island using small cut-outs of larger images. The length of the lines per image area was determined with the same routine as used for Fig. \ref{fig:Pt}(g). The Si coverage in percent of a monolayer is given in Fig. \ref{fig:BN}(d), assuming lines with a width of  one Si atom and an inter Si-separation of 2.77\,\AA\ (see below). As there is no rise in the coverage closer to the rim, we deduce that the Si does not intercalate from the sides of the island, but diffuses through the h-BN. This might be due to a strong binding of the h-BN edges to the metal due to the open bonds of the h-BN flakes. This is in line with DFT calculations for h-BN on Ir(111) \cite{Hagen2016} revealing a relatively strong binding of the h-BN edges to the metal substrate.

The Si does also not intercalate via grain boundaries as evidenced by Fig. \ref{fig:BN}(e), showing a fully closed h-BN layer on top of Pt(111) after Si deposition. The image does not show any accumulation of Si lines or clusters along the grain boundary. The lines even simply cross the boundary between the two differently oriented moir\'{e} structures, showing that the lines are barely influenced in their growth by grain boundaries of the h-BN.

Next, we also tried to lift the Si lines below the h-BN. We found that a way to lift them is heating to 200$^\circ$C, which causes all lines to vanish, while the $\sim$200\,pm high clusters grow in size (not shown). However, no growth of extended sub-monolayer high areas was observed, so that the mechanism is different to the one observed by exposing the bare Pt surface with Si chains to residual gas. The growth of the islands via the material from the lines can be used to estimate the atomic width of the lines. The coverage with clusters below the h-BN before heating was determined to be $3.1\pm0.5\%$ by flooding the image above the half-height of the clusters. The error was determined from varying the flooding limit in the range of the error on the visually determined cluster height. After heating to 200$^\circ$C, the clusters cover $6.7\pm0.7\%$ of the surface below the h-BN. The difference of $3.6\%$ corresponds nicely to the coverage of Si atoms in the lines of $3.5\pm0.7\%$ prior to annealing (Fig. \ref{fig:BN}(d)), if one  assumes a width of one atom and an inter-Si-distance equal to the Pt-Pt distance in the Pt(111) surface. In turn, division of these two values gives a line width of $1.0\pm0.3$ atoms. Thus, the embedded lines below the h-BN appear to be atomically thin one-dimensional Si chains. Such an estimate of the width could not be realized on the bare Pt(111) as we were not able to deduce a clear atomic model for the more compact, embedded areas (likely a precursor of the surface silicide \cite{Svec2014}). Thus, we could not estimate the amount of atoms contained within them.

While the line network below the h-BN exhibits similarities to the one on bare Pt(111), there is one striking difference. While on bare Pt the lines are oriented preferentially along the $\langle \overline{1}\overline{1} 2\rangle$-direction, beneath the h-BN they are preferentially orientated along the three $\langle 1\overline{1}0\rangle$-directions. Figure \ref{fig:BN}(f) shows a histogram of the angle of the lines with respect to the atomic rows (inset in Fig.\ref{fig:Pt}(a)). While it exhibits more variations than Fig. \ref{fig:Pt}(f), in line with the stronger meandering of the lines in Fig. \ref{fig:BN} than in Fig. \ref{fig:Pt}, its maxima are along the $\langle 1 \overline{1} 0 \rangle$ directions, distinct from the case on the bare Pt(111). This will be tentatively explained by the influence of the strain within the h-BN layer.

\begin{figure}
\includegraphics[width=0.45\textwidth]{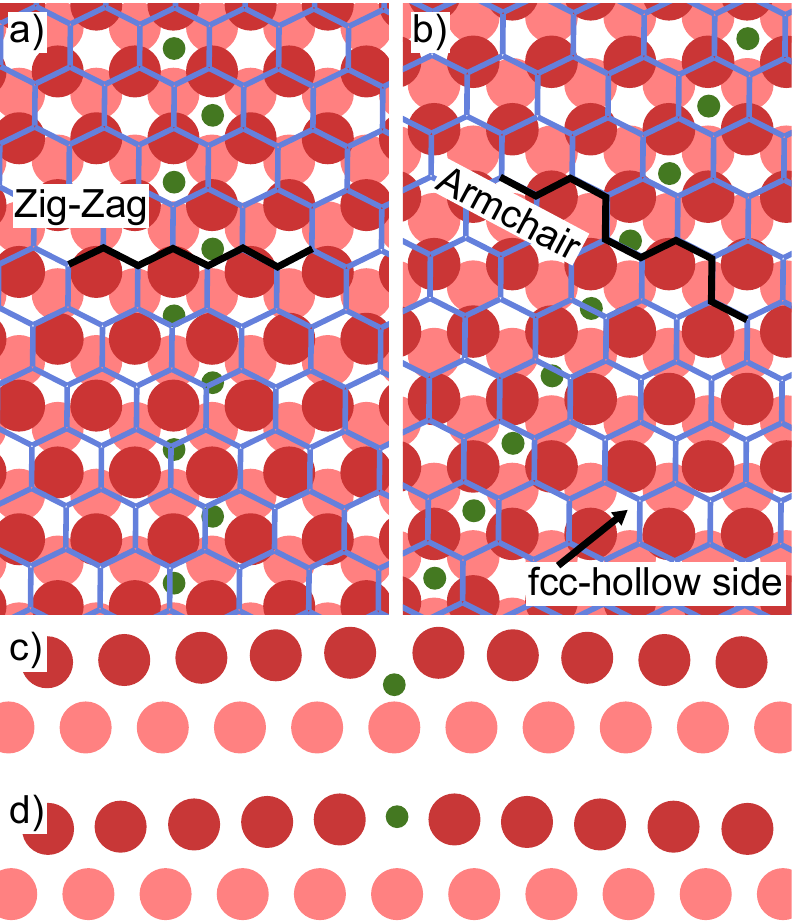}
\caption{\label{fig:Mod} (Color online) Model of the 1D Si chain embedded in Pt(111) below the h-BN. The dark red circles represent the atoms of the top Pt layer, the light red ones are those of the layer below. Green circles represent the silicon atoms. The size of the circles represents the atoms' ionic radius \cite{Shannon1976}. Blue lines depict the h-BN lattice and black lines emphasize the direction in which this lattice is stretched due to the Si chains. (a) A chain in $\left[ \overline{1} \overline{1} 2 \right]$-direction stretching the h-BN in the zigzag-direction. (b) A chain in  $\left[ 0 \overline{1} 1 \right]$-direction stretching the h-BN in the armchair-direction. (c) Side-view of the model with the Si atom sitting in between the two top layers of Pt(111). (d) Side-view of the model with the Si atom siting inside the top layer of Pt(111).}
\end{figure}

We, thus now provide tentative atomic models (Fig. \ref{fig:Mod}) that might be the starting point to explain our intriguing experimental observations in detail. 
As discussed in the introduction, a Pt surface gains energy by incorporating further atoms. For platinum silicides, whose bare existence testifies to the solubility of Si in Pt, different bonding configurations are found. They range from more covalent Pt$_2$Si to more metallic PtSi \cite{Klepeis2001}. One has also observed an embedding of the silicon in the top layer at intermediate temperature \cite{Svec2014}, which is dubbed a surface Pt silicide. 

The atomic configuration of the bare Pt(111) surface offers most space for embedment in the hollow-site, fcc-type positions (Fig. \ref{fig:Mod}(b)). We propose that the silicon atoms are located there. Depending on the position of the next nearest Si neighbors, the result is a chain in $\langle \overline{1} \overline{1} 2 \rangle$ or $\langle 1 \overline{1} 0 \rangle$-direction. The resulting structures, with the known Si density (discussion above) are shown in Figs. \ref{fig:Mod}(a) and (b). In these structures, neighboring Si atoms have a distance of 277\,pm. This is much larger than in any bonding configuration of pure silicon, so one can assume that the chains are not the result of Si-Si interactions, but rather governed by the Si-Pt interactions. Assuming additionally that the main driving force for the Si embedment is the increase of electron density in the Pt surface \cite{Bott1993,Hohage1995,Sandy1992,Lundgren1999,Grutter1994,Holst1998,Zhang1999}, Si atoms should be placed at positions that barely dusturb the Pt lattice. Two such positions, depicting the ionic radius of the different atoms via its symbol size, are shown in Figs. \ref{fig:Mod}(c) and (d). Either the Si atom is located in the surface (Fig. \ref{fig:Mod}(d)), as observed for the surface Pt silicide \cite{Svec2014}, or it is in the largest empty space in between the two top layers of Pt(111) (Fig. \ref{fig:Mod} (c)), where the Si atom would be six-fold coordinated similar to PtSi \cite{Klepeis2001}. Without further calculations, we cannot decide which position is more favorable for the embedded Si chains.

However, we want to provide a model idea, why the orientation of the chains on bare Pt(111) and on h-BN covered Pt(111) is different. Inspecting the two different chain directions with h-BN on top (Figs. \ref{fig:Mod} (a),(b)), one notices that one configuration stretches the h-BN along the zig-zag-direction, while the other one stretches it along the armchair-direction (black lines of the h-BN lattice). Classical molecular dynamics calculations (employing finite-size h-BN layers) indeed show a difference in the Young's modulus $E$ for different stretching directions \cite{Thomas2016,Thomas2017}. Specifically, $E$ is smaller in the armchair-direction (755\,GPa at 300\,K) than in the zig-zag direction (769\,GPa at 300\,K). This might lead the chains to be preferentially orientated perpendicular to the armchair direction, when there is a h-BN layer on top. 

While it remains unclear, why the chains are orientated along the Pt-$\langle \overline{1} \overline{1} 2 \rangle$-directions without h-BN on top, we notice that dislocation lines in the bare Pt(111) surface induced by excess Pt or Co are orientated along that axis, too \cite{Bott1993,Hohage1995,Sandy1992,Lundgren1999,Grutter1994,Holst1998,Zhang1999}. 

\section{\label{sec:Con}Conclusion}

In conclusion, we have demonstrated that Si deposition onto Pt(111) at 300\,K forms a network of atomically thin 1D Si chains embedded into the Pt(111) surface. The chains disappear in UHV within hours due to residual gas adsorption, most importantly by CO. The chain network can be protected by a monolayer of h-BN, covering the Pt(111) surface before deposition of the silicon. This makes the chains stable for weeks in UHV and even stable in air for, at least, 10 min. The h-BN coverage, additionally, results in a rotation of the chains by 30$^\circ$, likely due to the anisotropic mechanical properties of h-BN.

The largely airtight protection of the one atom thick Si structures by a h-BN monolayer underlines the potential of 2D materials for protection of fragile heterogeneous atomic assemblies. This might be exploited for other self-organized structures, too.

\begin{acknowledgments}
We thank B.Voigtl\"{a}nder (FZ J\"{u}lich) for providing us with a Si evaporator \cite{MYSLIVECEK2002193} and A. Gr\"{u}neis (Univ. of Cologne) for help regarding the handling of borazine.
Most of the STM data was analyzed using the WSxM software \cite{WSxM}. In particular, the "Flatten plus"-function proved extremely useful. 
Funding by the Graphene Flagship (Contract No. 696656) and the Excellence Initiative of the German federal and state governments is gratefully acknowledged. 
P. Nemes-Incze acknowledges support form the Hungarian Academy of Sciences, within the framework of the MTA EK Lend\"{u}let Topology in Nanomaterials Research Group, grant no: LP2017-9/2017.
The computations were performed at the high-performance computer JURECA operated by the JSC at the Forschungszentrum J\"{u}lich. V. Caciuc and N. Atodiresei acknowledge financial support from the DFG through the Collaborative Research Center SFB 1238 (Project No. C01). 

\end{acknowledgments}

\bibliography{Paper_Silicon1Dchains}

\end{document}